\documentclass{article}
\usepackage{amssymb,graphicx,amsmath,axodraw}

\newcommand{\preprintno}[1]
{\vspace{-2cm}{\normalsize\begin{flushright}#1\end{flushright}}\vspace{1cm}}

\title{\preprintno{{\bf ULB-TH/02-20.}}CP Violation from Dimensional Reduction: Examples in 4+1 Dimensions.}
\author{N. Cosme\thanks{ncosme@ulb.ac.be}, J.-M. Fr\`ere\thanks{frere@ulb.ac.be}, 
L. Lopez Honorez\thanks{llopezho@ulb.ac.be}. \\
\textit{\normalsize{Service de Physique Th\'eorique, CP225}
Universit\'e Libre de Bruxelles,
Bld du Triomphe, 1050 Brussels, Belgium.}}

\begin{document}

\date{July - 2002.}
\maketitle

\begin{abstract}
We provide simple examples of the generation of complex mass terms and hence $CP$ violation
through dimensional reduction.
\end{abstract}

%%%%%%%%%%%%%%%%%%%%%%%%%%%%%%%%%%%%%%%%%%%%%%%%%%%%%%%%%%%%%%%%%%%%%%%%%%%%%%%%%%%%%%%%%%%%%%%%%%%%%%%%%%%%%%%%%%%
%%%%%%%%%%%%%%%%%%%%%%%%%%%%%%%%%%%%%%%%%%%%%%%% INTRO %%%%%%%%%%%%%%%%%%%%%%%%%%%%%%%%%%%%%%%%%%%%%%%%%%%%%%%%%%%%
%%%%%%%%%%%%%%%%%%%%%%%%%%%%%%%%%%%%%%%%%%%%%%%%%%%%%%%%%%%%%%%%%%%%%%%%%%%%%%%%%%%%%%%%%%%%%%%%%%%%%%%%%%%%%%%%%%%

\section{Introduction.}

We regard the $CP$ symmetry as fundamental. It is indeed the natural symmetry of gauge interactions, where it can
be traced directly to the unitary nature of the gauge groups. A "pure" gauge theory (that is, without scalar 
interactions, such as Yukawa terms or even masses) is indeed $CP$-invariant in $3+1$ dimensions at the classical level.
Although it might be argued that an alternate  source of violation is  to be found in quantum  anomalies 
(the so-called $\theta$ term), it is well-known that this effect can be rotated
away for massless fermions. 

In the Standard Model $CP$ violation thus comes only from the completely arbitrary Yukawa couplings.
The purpose of the quest for unification is  eventually to eliminate the need for such arbitrary coupling parameters,
possibly relating them to the gauge interactions. It is thus to be expected that a unified theory should be 
$CP$-invariant. In such a case, a breaking mechanism is needed. It could be spontaneous (non-alignment of condensates 
or scalar vev), but the possibility also exists to relate it to some dynamical effect, for instance here, to 
dimensional reduction \cite{thirring},\cite{autres}.

In section \ref{CP}, we will briefly discuss the discrete symmetries in $(d-1)+1$ dimensions, then construct
in section \ref{vector} an explicit example. This example, while providing the equivalent of an electric dipole
moment still has a rather fundamental problem, namely that the theory stays (up to a phase) vectorlike. Ways out are
presented in section \ref{chiral}.

%%%%%%%%%%%%%%%%%%%%%%%%%%%%%%%%%%%%%%%%%%%%%%%%%%%%%%%%%%%%%%%%%%%%%%%%%%%%%%%%%%%%%%%%%%%%%%%%%%%%%%%%%%%%%%%%%%
%%%%%%%%%%%%%%%%%%%%%%%%%%%%%%%%% P CP AND CPT IN d-1+1 %%%%%%%%%%%%%%%%%%%%%%%%%%%%%%%%%%%%%%%%%%%%%%%%%%%%%%%%%%
%%%%%%%%%%%%%%%%%%%%%%%%%%%%%%%%%%%%%%%%%%%%%%%%%%%%%%%%%%%%%%%%%%%%%%%%%%%%%%%%%%%%%%%%%%%%%%%%%%%%%%%%%%%%%%%%%%

\section{$P$, $CP$ and $CPT$ in $(d-1)+1$.} \label{CP}

While this issue has been extensively tackled before, we recapitulate here a few salient points, and try to
dissipate an apparent paradox \cite{Gavela}.

There is a possible ambiguity in the definition of $P$. In $3$ spatial dimensions, two definitions, namely the central
inversion $\overrightarrow{x} \rightarrow -\overrightarrow{x}$ and the specular reflexion, say $x_1 \rightarrow - x_1$, 
are equivalent modulo one spatial rotation.
For even $(d-1)$ spatial dimensions however, the specular reflexion stays a discrete symmetry, while the central inversion is simply
an element of the rotation group (with $det=-1$ and $+1$ respectively). Which is the best generalisation?

It turns out that the specular reflexion leads to the generalisation of the $CPT$ theorem, which is a strong reason 
to choose it. In all dimensions, $P$ may thus now be identified to $x_{d-1} \rightarrow - x_{d-1}$, or 
equivalently $x_1 \rightarrow - x_1$ \cite{Casadio}. 

Another well-known statement is that the coupling $\bar{\psi}\psi$
in $4+1$ dimensions is $P$-violating. This may seem paradoxical. Indeed a scalar term in $3+1$ dimensions can be viewed as
$P$-conserving, and we have just seen that $P$ can be defined in a universal way whatever the number of dimensions.
A clarification may be found in the more usual $3+1$ situation. Here indeed, we can have both
$$\bar{\psi}\psi= \bar{\psi}_L \psi_R + \bar{\psi}_R \psi_L$$
and
$$\bar{\psi} i \gamma_5 \psi= i (\bar{\psi}_L \psi_R - \bar{\psi}_R \psi_L).$$
It is easy in $3+1$ dimensions to go from one type of coupling to the other by a mere sign flip of the semi-spinor (say
$\psi_L \rightarrow -i \psi_L$). In fact, to achieve $P$ violation in $3+1$ dimensions through spin-$0$ couplings, the 
simultaneous presence of $\bar{\psi}\psi$ and $\bar{\psi} \gamma_5 \psi$ terms is needed. In $4+1$ dimensions, the
$\bar{\psi} \gamma_4 \psi$ component of the vector is automatically present in the kinetic part, and precisely corresponds 
to $i \bar{\psi} \gamma_5 \psi$ in $3+1$ language through $\gamma_4= i \gamma_5$. This in fact "locks" the definition and 
results in $\bar{\psi} \psi$ to be $P$-violating.

As we will see below, this is also at the origin of $CP$ violation in the dimensional reduction process. With the 
definition of $CP$ given above, the term $M\bar{\psi}\psi$ is easily seen to preserve $CP$ in $4+1$ dimensions while 
breaking both $C$ and $P$. For easy reference, we give below one possible representation of the $C$ and $P$ operators
in $4+1$ dimensions:

$$C^{-1}_{(4+1)} \gamma_A C_{(4+1)} = \gamma_A^t  \longrightarrow C= \gamma^1 \gamma^3 = \gamma^2 \gamma^0 \gamma^5,$$
$$ \psi^{P_{4+1}}( x_0,x_\mu,x_4) = \gamma^4 \psi( x_0,x_\mu,-x_4),$$
($A,B=0,1,2,3,4$).

%%%%%%%%%%%%%%%%%%%%%%%%%%%%%%%%%%%%%%%%%%%%%%%%%%%%%%%%%%%%%%%%%%%%%%%%%%%%%%%%%%%%%%%%%%%%%%%%%%%%%%%%%%%%%%%%%%%%%%
%%%%%%%%%%%%%%%%%%%%%%%%%%%%%%%%%%%%%%%%%%%%%%%%%%%%FIRST EXAMPLES%%%%%%%%%%%%%%%%%%%%%%%%%%%%%%%%%%%%%%%%%%%%%%%%%%%%
%%%%%%%%%%%%%%%%%%%%%%%%%%%%%%%%%%%%%%%%%%%%%%%%%%%%%%%%%%%%%%%%%%%%%%%%%%%%%%%%%%%%%%%%%%%%%%%%%%%%%%%%%%%%%%%%%%%%%%

\section{First Examples.} \label{vector}

We work for the moment in $d=4+1$ dimensions (the extension to $2n+1$ dimensions is easy) without specifying yet the nature 
of the extra spatial dimension (orbifold, compact or just infinite). Starting from a Lagrangian with $U(1)$ gauged  for the 
fermions:

$$\mathcal{L}= i\bar{\psi}D \!\!\!\!/\psi -M \, \bar{\psi} \psi,$$
with $D_B= \partial_B -i e A_B$, we observe immediately that breaking dimensional reduction from  $4+1$ into $3+1$ 
will introduce effective complex mass terms into the Dirac equation via non-vanishings contribution arising 
from $\partial_4$ or possibly the $A_4$ terms, which we denote generally by $X_4$, resulting in a mass structure:
$$\bar{\psi}(M+i \gamma_5 X_4) \psi.$$
Such a structure could lead to $CP$ violation (for instance, if a strong anomaly, linked to a second gauge group is also present,
or, as was studied by Thirring \cite{thirring}, in the case of a non-minimal coupling of the photon).
In the case of a pure minimal-coupling 
U(1) theory, the complex mass term can however be rotated away by a chiral rotation in $3+1$ dimensions, and $CP$ 
violation thus requires at least an extension of the gauge group.

Before moving to such extensions, let us pause now to consider the possible origins of the contribution $X_4$.

The case considered by Thirring is the simplest; if the $4$th dimension is compactified, and $X_4$ simply corresponds
to the Kaluza-Klein mass $\frac{n}{R}$. The result is a tower of states, with no $CP$-violating effect for 
the fundamental. This offers little hope to relate to the observed phenomenology.

If we want to separate the $CP$ violation from the use of the excited Kaluza-Klein states, and thus bring it into the realm
of the (observed) zero-mode particles, or use an altogether different dimensional 
reduction scheme, the obvious solution is to assume some vacuum expectation value for the $4$th component of the gauge field itself:
$$\langle A_4(x,y,t) \rangle \neq 0,$$
(for now on, $x$ stands for the usual spatial coordinates and $x_4=y$).

Clearly such a statement is not gauge invariant as such, since the value of $A_4$ at given $y$ can always be rotated away. The
corresponding gauge invariant quantity is the line integral of $A_4$ over a suitable path:
$$\int dl \; A_4 .$$
Since we want to keep Lorentz invariance of the remaining $3+1$ dimensions, we will take this path entirely in the $y$
direction and write:
$$ X_4= \int dy \; A_4 ,$$
allowing $X_4$ to be time and x independent.

The upper and lower bounds of  this integral may vary according to the dimensional reduction scheme: from $-\infty$ to 
$+\infty$ for non-compactified $y$ (including the case of localisation on a defect), on a circle $[0, 2\pi R]$ for the 
Kaluza-Klein scheme, on a segment like $[0, \pi R]$ for an orbifold approach. 
In the case of a closed loop, this is just the usual Wilson loop
contribution, and can be thought of  as the flux of $ \overrightarrow{\nabla}\times \overrightarrow{A}$ through the 
(however unphysical) cross section of the torus. 

Alternatively \cite{Hall}, in the case of an orbifold, $A_4(y)$ can be 
gauged away, resulting in an equivalent formulation with non-periodical boundary conditions:
$$\psi'(y)= e^{-i \int_0^y dy A_4(y) } \psi(y).$$

The use of such a line integral to break down symmetry has been developed in details by Hosotani in the framework of 
dynamical symmetry breaking \cite{Hosotani}. We will not discuss here the mechanism for generating such 
a vacuum expectation value, which amounts as seen to a kind of boundary condition;
 we turn instead to the physical realisation of $CP$ violation.

\medskip
In the case of a pure $U(1)$, we have already mentioned that the phase appearing in the mass matrix can be safely 
rotated away. This was not the case in the model discussed by Thirring. Here indeed, the $U(1)$ Yang-Mills field in 
fact originates from the $g_{4\mu}$ components of the metric tensor, and inclusion of a torsion term in the coupling 
to fermions results in a non-minimal coupling:
$$\kappa F^{\mu\nu} \bar{\psi} \sigma_{\mu\nu} \psi,$$
with $F^{\mu\nu}= \partial^\mu A^\nu - \partial^\nu A^\mu$.
If $(M+i\gamma_5 X_4)$ and $\kappa$ are not simultaneously real, rotating the phase in the mass term brings an 
imaginary component in the magnetic coupling:
$$\kappa' F^{\mu\nu} \bar{\psi} i\sigma_{\mu\nu} \gamma_5 \psi,$$
which corresponds in fact to an electric dipole moment, clearly a $CP$-violating observable.

\medskip
While we prefer to avoid such non-minimal couplings (which are both non-renormalisable in 3+1 and in contradiction with observation),
a similar situation would occur if we have simultaneously
minimally coupled $U(1)$ and $SU(3)$ terms (like in strong interactions); this time the sum of the phase 
in the mass term  and of the $\theta$ term (corresponding, in the reduced dimensions, to the anomaly $\theta \;
\tilde{G}^{\mu\nu} G_{\mu\nu}$) induces a $CP$ violation.

\medskip

Of more interest to us however, for later generalisation, is a simple extension based on the $SU(2)$ group that we 
propose here. Here indeed, neither non-minimal coupling, nor non-perturbative effects are needed.

We start from the Lagrangian:
$$\bar{\Psi}i (\partial^A -i W^A_a \tau^a) \gamma_A \Psi + M \bar{\Psi}\Psi,$$
and assume both $M\neq 0$ and
$\langle W_4 \rangle =\int dy \; W_4(y) =\left(\begin{array}{cc} w&  \\  &-w \end{array}\right)$. 
This results in the effective $3+1$ Lagrangian:
$$\left(\begin{array}{cc}\bar{\psi_1} & \bar{\psi_2}\end{array}\right)i(\partial^\mu -i W^\mu_a \tau^a) \gamma_\mu
 \left(\begin{array}{c}\psi_1 \\ \psi_2\end{array}\right) 
+ \left(\begin{array}{cc}\bar{\psi_1} & \bar{\psi_2}\end{array}\right)\mathcal{M}
 \left(\begin{array}{c}\psi_1 \\ \psi_2\end{array}\right),$$ 
where 
$$\mathcal{M}= \left(\begin{array}{cc} M+iw\gamma_5&  \\  &M-iw\gamma_5\end{array}\right).$$
The mass matrix can be diagonalised generally by a bi-unitary transformation, $\mathcal{M}'= U^\dagger_R \mathcal{M} U_L$.
In fact, in the present case, the problem is partially undetermined and we can choose ($\alpha=\gamma_5 \arctan{\frac{w}{M}}$):
$$U_R=\Bbb{I}, \qquad U_L=\left(\begin{array}{cc} e^{-i\alpha}&  \\  &e^{i\alpha} \end{array}\right).$$
With the fermion masses now diagonal (and degenerate), we obtain two massive $W^\pm$ and 
one massless $W^3$ gauge bosons, with the breaking of the gauge symmetry according to the Hosotani mechanism and the 
effective Lagrangian. The coupling of $W^+$ and $W^-$ is no longer purely vectorial, but includes a phase between the $L$ 
and $R$ parts. As a result, a "$W^3$-dipole moment" is induced at one loop level (see Figure 1 for one example 
of a contribution).
\begin{center}
\begin{figure}[t]    
   \begin{picture}(200,110)(0,0) 
      % EDM
      \ArrowLine(185,60)(215,60) 
      \Text(200,80)[]{$e^{2i\alpha}$}
      \Vertex(215,60){2} 
      \Line(215,60)(305,60)
      \Text(215,45)[]{$L$}
      \Vertex(305,60){2} 
      \Text(305,45)[]{$R$}
      \Text(285,60)[]{$\times$}
        \Text(285,50)[]{$m$}
      \ArrowLine(305,60)(335,60)
      \PhotonArc(260,60)(45,0,180) {-4}{8}
        \ArrowArcn(260,60)(35,120,60)
      \Text(260,110)[b]{$W_{\pm}$} 
      \Vertex(260,60){2} 
      \Photon(260,0)(260,60){-4}{5} 
      \Text(280,15)[l]{$W_3$}
    \end{picture} 
        \caption{} 
    \label{fig:fig1}     
\end{figure}  
\end{center}

The interest of this model comes for 3 reasons:
\begin{itemize}
\item $CP$ violation, dimensional reduction and breaking of the internal symmetry are intimately linked,

\item the approach is purely perturbative,

\item the $CP$ violtation appears in a Kobayashi-Maskawa-like matrix.
\end{itemize}
Obviously this toy model has also strong limitations, which we list briefly:
\begin{itemize}
\item The remaining massless boson cannot be identified with the photon:  this can easily be solved by extending to
$SU(2)\times U(1)$ (but then the breaking pattern is still a problem, as breaking in the triplet leaves a massless
$Z$ boson).
\item The coupling of $W^+$ and $W^-$ is not chiral. While not purely vectorlike in the mass eigenstate basis
($L$ and $R$ have different phases), it is purely vectorlike in the current basis, and certainly does not differentiate
significantly between $L$ and $R$. 
This is particularly bothersome since the $CP$ generation mechanism appears in the same
time here to be directly linked to the presence of both $L$ and $R$ couplings to the gauge bosons (and of the $W^3_4$ in particular).
\end{itemize}
We will not address the  mass patterns here, and thus will not respond further to the first issue. 
We deal instead with the second issue in the next section. Let us announce the strategy:
\begin{itemize}
\item Obviously in $4+1$ dimensions, vectorlike couplings are automatic due to the Lorentz structure.
\item Localisation on a defect (domain wall) or equivalently an orbifold formulation will eliminate either $L$ or $R$ 
components.
\item We choose the defect structure to be part of the internal group; as a result the symmetry group is broken 
(outside of the defect) and, according to their couplings only some $L$ or $R$ components are localised.
\item The Hosotani terms then links these remaining $L$ and $R$ components.
\end{itemize}
Explicit examples based on $SU(3)$ and $SU(4)$ are discussed in the next section.

%%%%%%%%%%%%%%%%%%%%%%%%%%%%%%%%%%%%%%%%%%%%%%%%%%%%%%%%%%%%%%%%%%%%%%%%%%%%%%%%%%%%%%%%%%%%%%%%%%%%%%%%%%%%%%%%%%%%%%%
%%%%%%%%%%%%%%%%%%%%%%%%%%%%%%%%%%%%%%%%%%%%%%%CHIRAL EXAMPLE%%%%%%%%%%%%%%%%%%%%%%%%%%%%%%%%%%%%%%%%%%%%%%%%%%%%%%%%%%
%%%%%%%%%%%%%%%%%%%%%%%%%%%%%%%%%%%%%%%%%%%%%%%%%%%%%%%%%%%%%%%%%%%%%%%%%%%%%%%%%%%%%%%%%%%%%%%%%%%%%%%%%%%%%%%%%%%%%%%

\section{Chiral Examples.} \label{chiral}

In order to differentiate clearly $L$ and $R$ gauge interactions in the reduced Lagrangian, we thus now consider
topological defects which are part of the internal group. As an example, and to be explicit, we will focus on a domain wall 
scalar field in the adjoint representation.

As it is well known \cite{Rubakov}, such a defect coupled to fermions localises in its core massless fermionic zero modes 
with a defined chirality related to the sign of the coupling. For an adjoint scalar coupled to fermions in the fundamental
representation, we choose our basis to write the breaking direction as a diagonal operator. As a consequence, the following zero
 modes are selected :
$$\left(\begin{array}{ccccc} \psi^1_L & \psi^2_L  &\hdots  &\psi^i_R   &\hdots \end{array}\right).$$

Subsequently, as we wish to provide masses to fermions, we have to add scalars which will
acquire a constant vev. Those scalars then break  the group both inside and outside the wall, at the difference of the domain 
wall field which breaks the group only outside the wall. 
Finally, to get complex masses, the Hosotani term has also to be in a non-diagonal direction. 
\medskip

Let now turn to an explicit $SU(3)$ model in order to illustrate the idea. As only $3$ chiral localised states arise for 
the fundamental representation, we expect to form at most one massive and one massless fermion, so even if this example 
shows both a complex mass term and chiral couplings, $CP$ violation will usually be avoided by rotating away the phase. 
Yet, this case gives the building principles, we will after that consider an $SU(4)$  which will provide all the desired 
features. 
\medskip
\medskip
%%%%%%%%%%%%%%%%%SU(3)%%%%%%%%%%%%%%%%%%%%%%

We start with the following $SU(3)$ invariant fermionic Lagrangian in $4+1$ dimensions:

$$\bar{\Psi}i (\partial^A -i W^A_a \lambda^a) \gamma_A \Psi,$$
to which we first add a domain wall $\Phi$ along the $4$th coordinate:
$m_\Phi \bar{\Psi}\Phi \Psi$.

The choice of $\Phi$ in the $\lambda_8$ direction of the $SU(3)$ algebra implies thus a confined fermionic representation
in the form:\footnote{Here, we must keep in mind that the zero modes are localised differently due to the different strength
of the coupling. The localisation of the gauge fields (not discussed here) must take this into account to maintain charge
universality.} $\left(\begin{array}{ccc} \psi^1_L & \psi^2_L & \psi^3_R\end{array}\right).$
Now, we can fill in the second diagonal direction of the group $\lambda_3$ with another scalar $\chi$ which, when acquiring
a vev, will break the group to $U(1) \times U(1)$. As it has already been said, this scalar cannot generate mass terms between 
zero modes, and we must therefore introduce a third scalar $H$, for instance in the $\lambda_4$ direction, which couples
$\psi^1_L$ to $\psi^3_R$ and breaks the group to $U(1)$. All couplings are real in $4+1$ dimensions. 
The Hosotani term is forced to be parallel to $H$, and generates the following mass term:

$$\bar{\psi}^1_L \frac{1}{2} ( m_H \langle H \rangle + i \gamma_5 w) \psi^3_R + h.c..$$
As announced, in this minimal set up, we only generate one fermion mass (whose phase can be removed).
Nevertheless, we proceed to write the reduced gauge interactions to find out they are indeed chiral.

Of course, the approach only makes sense if the remaining (in particular) massless gauge fields are localized on the defect; this is however a common problem
to such defect-based models \cite{Dvali}, and will not be tackled here.

Now, taking the localisation of fermions into account, we can see that the only $W_1$, $W_2$, $W_3$ and $W_8$ gauge bosons
are involved in the effective theory, that is in interactions between zero modes. Indeed, the other ones provide  
interactions between, for instance, a localised left-handed zero mode and (the L part of) an unlocalised fermion with mass  
of the order of the confining scale. Such interactions therefore are not observable, and do not belong to the effective Lagrangian.
The initial gauge interaction eventually reduces thus to the effective interactions:
\begin{itemize}
\item  charged  and neutral currents:
$$ \frac{1}{\sqrt{2}} W^+_\mu  \bar{\psi}^1_L    \gamma^\mu \psi^2_L, \qquad
 \frac{1}{2} Z_\mu \bar{\psi}^1_L   \gamma^\mu \psi^1_L, \qquad
-\frac{1}{2} Z_\mu \bar{\psi}^3_R     \gamma^\mu \psi^3_R;$$
\item and the remaining $U(1)$ current:
$$\frac{-1}{2\sqrt{3}} A_\mu \bar{\psi}^1_L   \gamma^\mu \psi^1_L, \qquad
 \frac{-1}{2\sqrt{3}}  A_\mu \bar{\psi}^3_R    \gamma^\mu \psi^3_R,\qquad
 \frac{1}{\sqrt{3}}  A_\mu \bar{\psi}^2_L     \gamma^\mu \psi^2_L;$$
\end{itemize}
where we have written the interactions in terms of the mass and $U(1)$ eigenstates.
As a result, we get indeed effective gauge interactions which are chiral and possess an electroweak-like structure, although
one member of the effective fermion doublet is still left massless.

%%%%%%%%%%%%%%%%%%SU(4)%%%%%%%%%%%%%%%%%%%%%%
\medskip
\medskip
Following the same path, we  now turn to $SU(4)$. Let us consider the vacuum configuration:
$$\Phi = \frac{\phi(y)}{2} \left(\begin{array}{cccc}
1 &  &  & \\ 
 & 1 &  & \\ 
 &  & -1 & \\
& & &-1
\end{array}\right),
 \qquad \chi = \frac{\langle \chi \rangle}{2} \left(\begin{array}{cccc}
0 &  &  & \\ 
 & 0 &  & \\ 
 &  & 1 & \\
 &  &  &-1
\end{array}\right),
 \qquad \eta = \frac{\langle \eta \rangle}{2} \left(\begin{array}{cccc}
1 &  &  & \\ 
 & -1 &  & \\ 
 &  & 0 & \\
 &  &  &0
\end{array}\right),$$
where $\Phi$ provides the domain wall.
As seen above, the selection of chiral zero modes for the different fermions of the quadruplet
reduces there the gauge interactions of the fermions to $SU(2)_L \times SU(2)_R \times U(1)_A$ group
between zero modes: $\left(\begin{array}{cccc} u^1_L & d^1_L  & u^2_R  & d^2_R \end{array}\right)$.
The scalar fields $\eta$ and $\chi$, acquire a constant vev, and  break down respectively the $SU(2)_L$ and the $SU(2)_R$ subgroups.
Those fields fill in all the diagonal space of the algebra. 
After that, the generation of fermion masses needs non-diagonal scalars $H^1$ 
and $H^2$; e.g.:
$$H^1 = \frac{\langle H^1 \rangle}{2} \left(\begin{array}{cccc}
 0& 0 & 1 & 0\\ 
 0& 0 & 0 &0 \\ 
 1& 0 & 0 &0 \\
0&0 &0 &0
\end{array}\right)=\langle H^1 \rangle \lambda_4 ,
 \qquad H^2 = \frac{\langle H^2 \rangle}{2} \left(\begin{array}{cccc}
0 & 0 & 0 &0 \\ 
0 & 0 & 0 &1 \\ 
0 & 0 & 0 & 0\\
 0&  1&  0&0
\end{array}\right)=\langle H^2 \rangle \lambda_{11}.$$

Since those two breakings commute, they clearly minimise their interaction potential, but moreover allow the 
Hosotani term to get a component in each direction without cost of energy:
$$\int dy\;  W^4 = w_4 \lambda_4 + w_{11} \lambda_{11}.$$
This feature provides actually two masses with two phases, i.e.:
$$ \bar{u}^1_L \;\;\frac{1}{2}(m_1 \langle H^1 \rangle +i w_4 \gamma_5) \;\;u^2_R + 
\bar{d}^1_L \;\;\frac{1}{2}(m_2 \langle H^2 \rangle +i w_{11} \gamma_5)\;\; d^2_R + h.c.. $$
It can be easily checked that both phases cannot be removed completely from the Lagrangian. Indeed, considering the 
effective gauge interactions (dropping the 1 and 2 fermionic index):

$$\mathcal{L}_{C.C.}= \frac{1}{\sqrt{2}} W^+_{\mu L} \;\; \bar{u}_L \gamma^\mu d_L 
+ \frac{1}{\sqrt{2}} W^+_{\mu R} \;\; \bar{u}_R \gamma^\mu d_R + h.c.,$$

$$\mathcal{L}_{N.C.}= \frac{1}{2} (Z_\mu + \frac{A_\mu}{\sqrt{2}}) \;\; \bar{u}_L \gamma^\mu u_L 
+\frac{1}{2} (-Z_\mu + \frac{A_\mu}{\sqrt{2}}) \;\;\bar{u}_R \gamma^\mu u_R
+ \frac{1}{2} (Z'_\mu - \frac{A_\mu}{\sqrt{2}}) \;\; \bar{d}_L \gamma^\mu d_L 
+\frac{1}{2} (-Z'_\mu - \frac{A_\mu}{\sqrt{2}}) \;\; \bar{d}_R \gamma^\mu d_R, $$
in terms of the mass and U(1) eigenstates,
we cannot transform the masses to get them real without  obtaining a combination of the two phases in the 
charged currents.

\bigskip\medskip

%%%%%%%%%%%%%%%%%STABILITY%%%%%%%%%%%%%%%%%%%%%

For the sake of completeness, we discuss briefly the stability of the potential. 
Since all the considered scalars are in the adjoint of the group, we can generally take an interaction
potential which is minimal when all fields are orthogonal, e.g.: $\rho\; Tr \Phi H + \xi\; (Tr \Phi H)^2.$
The Hosotani term implies an effective potential for gauged scalars from the last component of
the covariant derivative, namely:
$(Tr[W_4,\Phi])^2$. This contribution is minimal for the two involved fields aligned or
commuting together. This term can be used to secure the orientation of one of the scalars parallel to the
Hosotani breaking direction. We have constructed explicit examples at the cost of small Yukawa couplings.
\bigskip\medskip

The model discussed here is not yet realistic in that charge assignations in the fundamental of $SU(4)$ are not compatible
with the observed ones (as could be expected, since SU(4) is not a suitable unification group). 
Also, to have $CP$ violation through the $W_L$ alone (here, the $W_R$ also participate), more generations are needed.

However, this example shows clearly that $CP$ violation can be generated in a fundamentally $CP$ symmetrical framework
through dimensional reduction.

\section*{Acknowledgments.}
We thank: P. Ball, Y. Gouverneur, N. Borghini, C. Boehm, M. Libanov, S. Troitsky for discussions. This work is 
supported in part by IISN, la Communaut\'{e} Fran\c{c}aise de Belgique (ARC), and the belgian federal government (IUAP).

%%%%%%%%%%%%%%%%%%%%%%%%%%%%%%%%%%%%%%%%%%%%%%%%%%%%%%%%%%%%%%%%%%%%%%%%%%%%%%%%%%%%%%%%%%%%%%%%%%%%%%%%%%%%%%%%%%%%%
%%%%%%%%%%%%%%%%%%%%%%%%%%%%%%%%%%%%%%%%%%%%%%BIBLIOGRAPHIE%%%%%%%%%%%%%%%%%%%%%%%%%%%%%%%%%%%%%%%%%%%%%%%%%%%%%%%%%%
%%%%%%%%%%%%%%%%%%%%%%%%%%%%%%%%%%%%%%%%%%%%%%%%%%%%%%%%%%%%%%%%%%%%%%%%%%%%%%%%%%%%%%%%%%%%%%%%%%%%%%%%%%%%%%%%%%%%%

\end{document}